\RequirePackage{fix-cm}
\documentclass{svjour3}                     
\smartqed  
\usepackage{graphicx}
\usepackage{aas_macros}
\usepackage{amsmath}
\usepackage{amssymb}
\usepackage[dvipsnames]{xcolor}

\newcommand{\HI}{H\,{\sc i}}
\newcommand{\CI}{C\,{\sc i}}

\begin{document}

\title{Molecular hydrogen in absorption at high redshifts.
}
\subtitle{Science cases for CUBES.}


\author{S.A.~Balashev$^1$ \and P.~Noterdaeme$^{2,3}$    
}


\institute{  $^1$ Ioffe Institute, {Politekhnicheskaya 26}, 194021 Saint Petersburg, Russia \\
              $^2$ Franco-Chilean Laboratory for Astronomy, Camino El Observatorio 1515, Las Condes, Santiago, Chile \\
              $^3$ Institut d'Astrophysique de Paris, CNRS-SU, UMR\,7095, 98bis bd Arago, 75014 Paris, France \\
              \email{s.balashev@gmail.com, noterdaeme@iap.fr}           
}

\date{Received: date / Accepted: date}

\maketitle

\begin{abstract}
Absorption lines from molecular hydrogen ($\rm H_2$) in the spectra of background sources are a powerful probe of the physical conditions in intervening cold neutral medium. 
At high redshift, $z>2$, $\rm H_2$ lines are conveniently shifted in the optical domain, allowing the use of ground-based telescopes to perform high-resolution spectroscopy, which is essential for a proper analysis of the cold gas. We describe  recent observational progress, 
based on the development of efficient pre-selection techniques 
in low-resolution spectroscopic surveys such as the Sloan Digital Sky Survey (SDSS). The next generation of spectrographs with high blue-throughput, 
such as CUBES, will certainly significantly boost the efficiency and outcome of follow-up observations. In this paper, we discuss high priority science cases for CUBES, building on recent $\rm H_2$ observations at high-z: probing the physical conditions in the cold phase of regular galaxies and outflowing gas from active galactic nucleus.
\keywords{Molecular hydrogen \and Quasar absorption lines \and Active galactic nuclei}
\end{abstract}

\section{Introduction}
\label{sec:intro}
Molecular hydrogen ($\rm H_2$) -- the most abundant molecule in the Universe -- traces the cold ($T\sim100$\,K) phase of the neutral gas. Since this phase is inclined to gravitational collapse through Jeans instability, understanding its physical and chemical properties is fundamental for a complete theory of star-formation throughout  the Universe’s history. Molecular hydrogen can be detected directly through electronic absorption lines in the Lyman and Werner bands ($\sim91-110$\,nm rest-frame) \cite{Carruthers1970,Spitzer1974}. At high redshift ($z>2$), $\rm H_2$ lines are conveniently shifted into optical domain and become accessible for the ground telescopes. Because of their brightness, quasars are generally used as background sources, although there is growing interest in the use of gamma-ray burst afterglows as well \cite{Prochaska2009,Kruhler2013,Bolmer2019}. 
The large collecting area of ground-based telescopes, combined with  
high-resolution spectrographs, permits to resolve narrow $\rm H_2$ lines that typically arise from several rotational levels in the cold $\sim 100$\,K neutral medium (see e.g. \cite{Balashev2019}). The ability to detect rotational levels is key to probe the physical conditions in the gas by modelling the relative population of H$_2$ levels (e.g. \cite{Klimenko2020}).

In this paper, we briefly describe an historical perspective and recent progress in studying diffuse molecular and translucent gas at high-$z$, through $\rm H_2$ absorption lines, mostly with Ultraviolet-Visual Echelle Spectrograph (UVES, \cite{Dekker2000}) and X-shooter \cite{Vernet2011} instruments on the Very Large Telescope (VLT) in Paranal. We discuss the tremendous boost in this field that would be enabled by Cassegrain U-Band Efficient Spectrograph (CUBES, \cite{CUBES}), in particular thanks to its efficient coverage of $\sim300-400$\,nm wavelengths. Key benefits will be: (i) an increase of the sample sizes of reachable systems by an order of magnitude. The corresponding systems are already available from pre-selection in the Sloan Digital Sky Survey (SDSS, \cite{York2000}) but other surveys will also contribute to discover new ones; (ii) the access to more extreme and elusive environments (e.g. more dusty/molecular rich). This will allow us not only to access the thermal state of the cold phase of the neutral medium of the overall population of high-$z$ galaxies but also to investigate the active galactic nuclei (AGN) feedback and quasar environment as probed by a recently discovered population of proximate $\rm H_2$ absorption systems\footnote{"Proximate" here refers to  absorption systems with redshift $z_{\rm abs}$ similar to that of the quasar emission, $z_{em}$, typically with $\Delta v \approx c \cdot |z_{\rm abs} - z_{\rm em}|/ (1+z_{\rm em})< \rm few \times 1000\,km\,s^{-1}$, where $c$ is the speed of light. Such absorption systems can be physically associated with the quasar and/or its host galaxy.}.

\begin{figure*}
\includegraphics[width=1.0\textwidth]{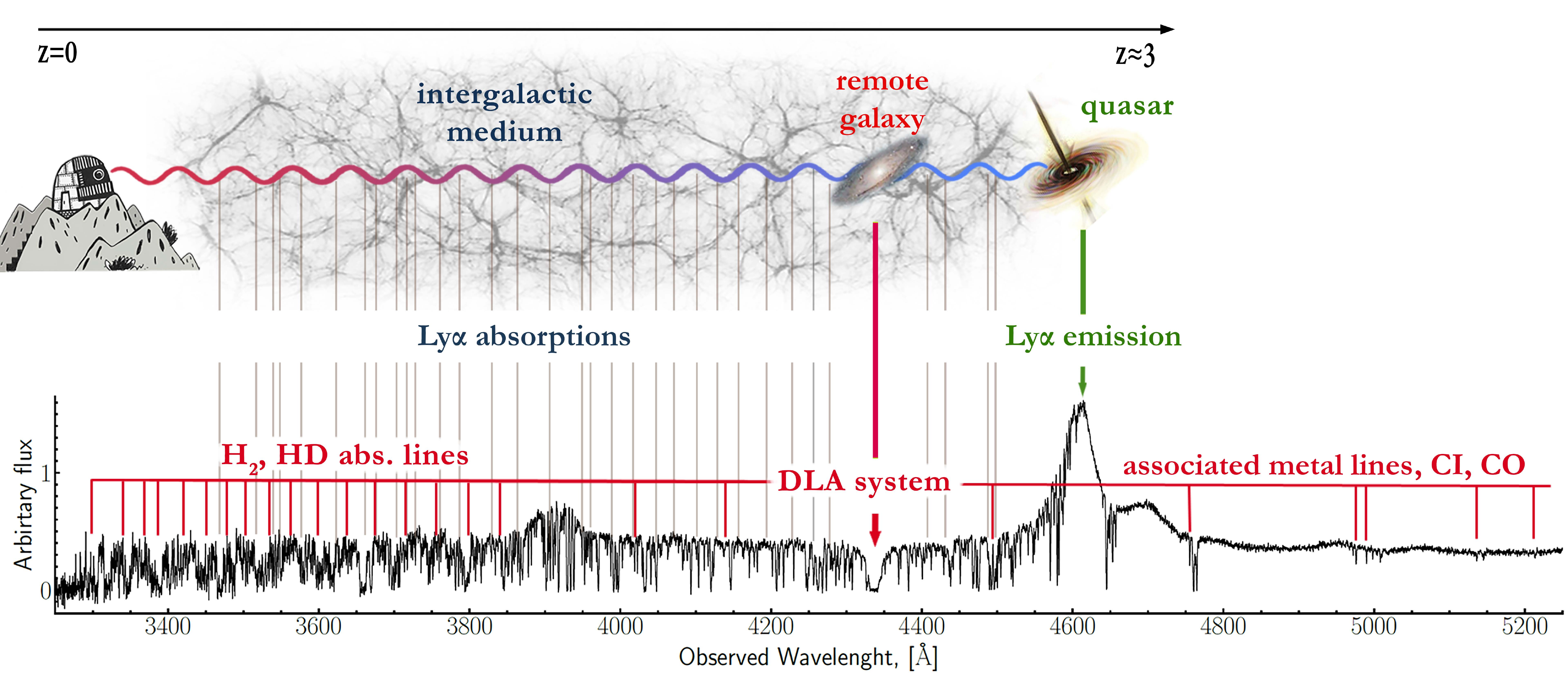}
\caption{Sketch of the studies of H$_2$-bearing damped Ly-$\alpha$ systems (DLAs) at high redshift towards quasars. The red lines mark absorption lines associated with DLAs, that are likely produced by interstellar or circum-galactic neutral gas located on the line of sight. These lines include \HI\ Lyman series, metal absorption lines as well as lines from H$_2$, HD and CO molecules.}
\label{fig:QSO_abs}
\end{figure*} 

\section{Observations of H$_2$ absorption}
\label{sec:high_z}
Molecular hydrogen being a homonuclear molecule, it has no dipole moment in the ground electronic state. Therefore, in emission, $\rm H_2$ can be observed only under specific conditions in the interstellar medium (ISM) -- mainly in photo-dissociated regions, which represent dense and warm gas, with temperatures $\gtrsim 500$\,K -- the typical distances between rotational energy levels of orto- or para-H$_2$. In turn, the electronic Lyman ($\rm B^{1}\Sigma^{+}_{u} - \rm X^{1}\Sigma^{+}_{g}$) and Werner ($\rm C^{1}\Pi_{u} - \rm X^{1}\Sigma^{+}_{g}$) bands transition of $\rm H_2$ are permitted. These transitions were first detected in space via rocket flight
observations 
by \cite{Carruthers1970} and then extensively analysed using the Copernicus satellite \cite{Spitzer1974}, mostly towards bright nearby stars. These, as well as following observations by the Far Ultraviolet Spectroscopic Explorer (FUSE, \cite{Sahnow1996}) satellite, concerned the local group \cite{Shull2000,Tumlinson2002}. In total, about 140 sightlines with H$_2$ detection where identified in FUSE data, through a recent reanalysis by \cite{Shull2021}. 

\subsection{High redshifts}
In principle, and somehow paradoxically, H$_2$ in remote galaxies is easier to observe than locally, owing to the 
Lyman and Werner bands of H$_2$ being redshifted into the optical domain. This permits the use 
of large optical telescopes to collect high-resolution spectra ($R\gtrsim50000$\footnote{Corresponding to full width half maximum (FWHM) of the instrument function, ${\rm FWHM} = c/R \approx 6$\,km\,s$^{-1}$, where $c$ is the speed of light.}) of the background source. In comparison,  
low-$z$ observations have been predominantly performed at $R\lesssim20000$. The gain in resolution for high-$z$ observations is an important advantage, since it allows to fully resolve the H$_2$ bands and to perform a proper de-blending from the Lyman-$\alpha$ forest as well as between H$_2$ velocity components. The instrumental profile is then also getting much closer to the intrinsic width of individual H$_2$ components that typically have Doppler parameters of only a few km\,s$^{-1}$ or less, allowing much more precise measurements. 
At high redshifts, quasars and gamma ray burst (GRB) afterglows can be used as bright background sources, albeit with some limiting capabilities in case of GRBs due their transient (a rapid decrease in luminosity) nature. 
In practice, H$_2$ absorption lines are always found to be associated with high column densities of neutral hydrogen, with the corresponding \HI\ Lyman series lines being located in the same wavelength region (see Fig.~\ref{fig:QSO_abs}) 
The overall measured \HI\ column densities\footnote{Here and after the column densities, $N$, are expressed in particles (atoms or molecules) per cm$^{2}$} throughout H$_2$-bearing systems are typically $\log N(\text{\HI}) > 19$  and possibly much above. H$_2$ absorption systems hence belong to either the so-called Damped Lyman-$\alpha$ systems (DLAs), with $\log N(\text{\HI}) \ge 20.3$), or to strong sub-DLAs with $19 \le \log N(\text{\HI}) < 20.3$. A third sub-class has recently been introduced for the highest column densities, with $\log N(\text{\HI}) > 21.7$ and dubbed extremely strong DLAs (ESDLAs, \cite{Noterdaeme2014}).

The first $\rm H_2$ absorption system at high $z$ was detected serendipitously by \cite{Levshakov1985} towards PKS\,$0528-250$ using only 2\,\AA\ resolution spectrum obtained by Anglo-Australian 3.9m telescope \cite{Morton1980}. The second high-$z$ H$_2$ absorption system was detected only a decade later \cite{Ge1997} 
using the Multiple Mirror Telescope, indicating a very slow progress in this field in pre-VLT/Keck era.

\subsection{VLT/Keck era}
A new epoch of high-z H$_2$ absorption studies began with the advent of high-resolution spectrographs, High Resolution Echelle Spectrometer (HIRES, \cite{Vogt1994}) and UVES on the 8-m class telescopes 
Keck and VLT, respectively. This resulted in a increase of the number of detected molecular hydrogen absorption systems, with the most important contribution arising from  
the VLT, thanks to dedicated and regular science programs to search for H$_2$ \cite{Petitjean2000,Ledoux2003,Srianand2005}. The UVES database for H$_2$ systems comprised thirteen H$_2$ systems \cite{Noterdaeme2008}. On Keck telescope, in spite of a large observational program focusing on DLAs ($\gtrsim 100$ DLA were studied at $z\gtrsim 2$ \cite{Prochaska2003b}), 
H$_2$ was reported in only a handful of systems \cite{Prochaska2003a,Jorgenson2009}, mostly due a 
wavelength coverage not optimised from H$_2$ during observations. The semi-blind surveys on these telescopes (and also Magellan, see \cite{Jorgenson2014}) indicated that the incidence rate of H$_2$ in DLAs is pretty low $\sim 10$\% for H$_2$ column densities above $10^{17}$\,cm$^{-2}$ \cite{Noterdaeme2008}. An independent estimate of the incidence rate in the overall population of DLAs, $\sim4.0\pm0.5(\rm stat.)\pm1.0(\rm syst.)$\% for saturated H$_2$ absorption systems ($\log N(\rm H_2)>18$) was later obtained from the detection of H$_2$ lines in the composite DLA spectrum \cite{Balashev2018} constructed with SDSS. 

\begin{figure*}
\includegraphics[width=1.0\textwidth]{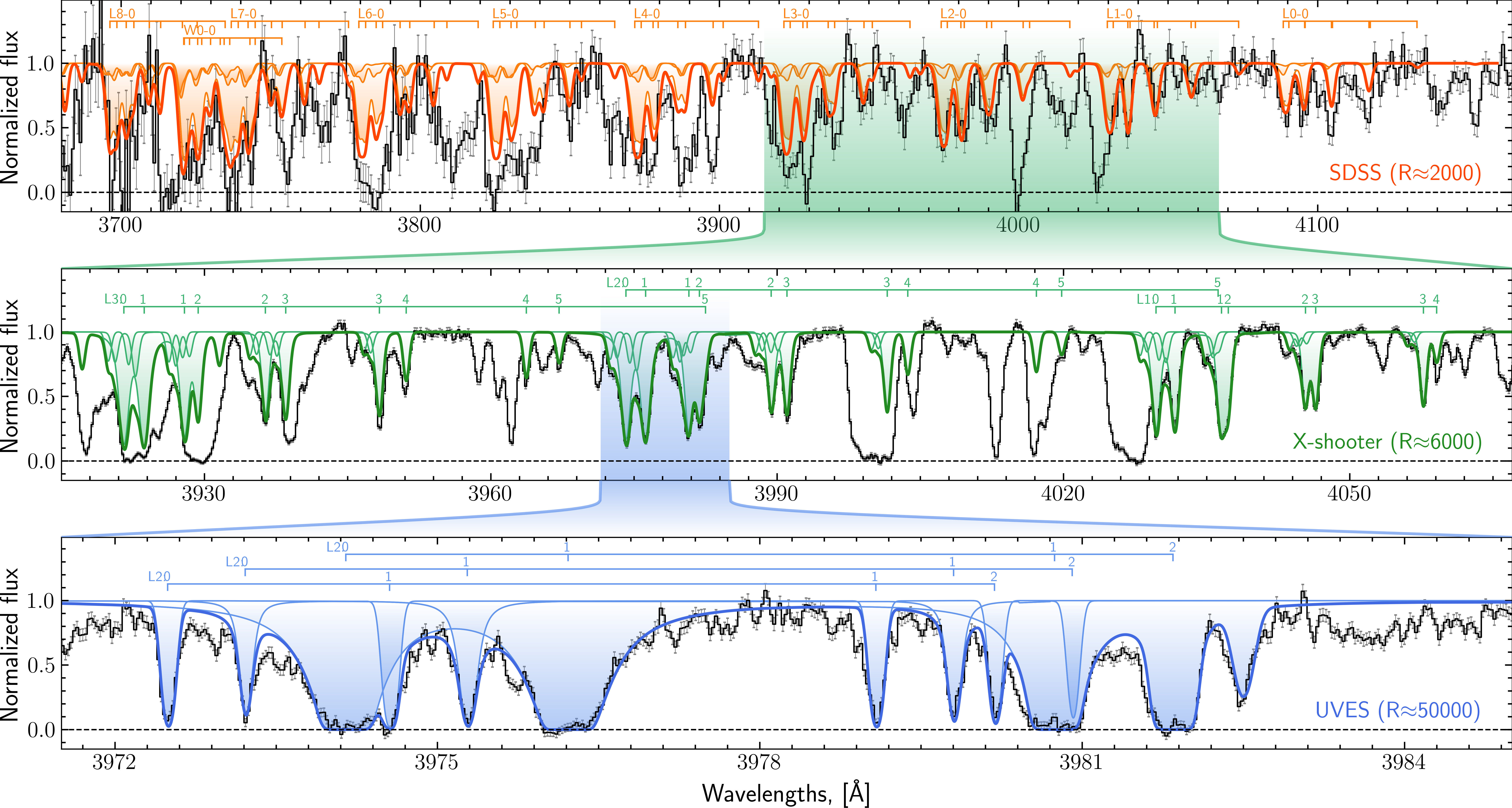}
\caption{Portions of spectra of J\,$1237+0647$ featuring H$_2$ absorption lines from a system at $z=2.625$, see \cite{Noterdaeme2010}. The top, middle and bottom panels correspond to the spectrum obtained using SDSS ($R\approx2000$), X-shooter ($R\approx6000$) and UVES ($R\approx50000$), respectively. The black and colored thick lines show the observed spectrum and total model profile of H$_2$ absorption system, respectively. The colored lines and regions in each panel show the profiles of individual velocity components, that are resolved and fitted using the UVES spectrum \cite{Noterdaeme2010}. The horizontal line with marks show the positions of H$_2$ absorption lines of the Lyman (denoted as $L$) and Werner (denoted as $W$) bands from different rotational levels, with the number above each tick indicate the rotational level.
}
\label{fig:J1237}
\end{figure*} 


\section{Pre-selection with SDSS}
\label{sec:sdss}
The low detection rate of H$_2$ among DLAs, together with the $\sim 15$\% incidence rate of high-$z$ DLA in quasar spectra results in a very low efficiency of blind searches, making this approach time-costly and impractical on highly over-subscribed telescopes.  
Fortunately, the availability of the SDSS, that collected several 10$^5$ quasar spectra with 
$R\approx 2000$, 
not only permitted the detection of many thousands high-$z$ DLAs/sub-DLAs  \cite{Noterdaeme2012,Parks2018}, but also triggered the development of efficient techniques for pre-selecting H$_2$-bearing systems. Indeed, it was shown that strong H$_2$ absorption systems (with $\log N(\rm H_2) \gtrsim 18$) can be either detected directly via their characteristic Lyman-Werner band signature in SDSS spectra (provided their quality is high enough and the Lyman-$\alpha$ forest not too dense) 
\cite{Balashev2014,Noterdaeme2019} or using \CI\ as a good tracer of H$_2$ \cite{Srianand2008,Noterdaeme2018}, in particular at high metallicity. Several C\,{\sc i} lines are generally located redwards of the Lyman-$\alpha$ emission, allowing their automated detection in SDSS \cite{Ledoux2015}. An example of the H$_2$-bearing DLA that has been initially found in SDSS and lately followed-up with both X-shooter and UVES is shown in Fig.~\ref{fig:J1237}. Finally, it was found that the incidence rate of H$_2$ significantly increases (up to $\sim 50$\%) in the ESDLA sub-sample from QSO \cite{Noterdaeme2015,Balashev2018,Ranjan2020} as well as GRB afterglows sightlines \cite{Bolmer2019}. This is most likely due to the corresponding lines of sight intercepting galaxies at small galacto-centric distances \cite{Arabsalmani2015,Lyman2017,Ranjan2018,Ranjan2020}, where the prevailing physical conditions, and in particular the gas pressure, favour the existence of the cold phase and the \HI/H$_2$ conversion \cite{Guimaraes2012,Noterdaeme2015,Balashev2017,Bolmer2019}.  

A comparative study of these three pre-selection techniques \cite{Balashev2019} indicates that they have different selection function in terms of the properties (such as metallicity, $N(\text{\HI})$, $N(\rm H_2)$, presence of CO, etc.) and therefore should complement each other. For example, regarding the metallicity, the CI-selected systems have on average an order of magnitude higher metallicities than ESDLA-selected ones, while direct search for H$_2$ lines pick up absorption systems in between \cite{Balashev2019}. It is also important to note, that these pre-selections were obtained based solely on SDSS owing to its statistical power. Current samples therefore inherit from the quasar selection functions of the different SDSS spectroscopic surveys \cite{Krogager2019}. In particular, optical selection tends to exclude highly reddened quasars and hence dusty and large H$_2$ column systems. 
This, together with the small cross-section of dense molecular clouds implies that the H$_2$ absorption systems currently probe mostly diffuse molecular gas (with exception of a few cases, see \cite{Noterdaeme2010,Balashev2017}).  

Notwithstanding, these techniques allowed us to significantly increase the number of confirmed H$_2$ absorption systems from $\sim 15$ to $\approx 50$ in the last decade, with an extremely efficient use of the observational time on the largest optical telescopes. This also permitted the study of systems spanning a very large range of properties of the underlying galaxy population. Among them, we can note the detection of: (i) Perseus-like cloud at $z\approx2$ \cite{Noterdaeme2017}; (ii) CO and 2175\AA\ dust bump in high-metallicity H$_2$-bearing DLAs \cite{Srianand2008,Noterdaeme2009,Noterdaeme2011}; (iii) large H$_2$ column density and low-metallicity absorbers, representing CO-dark molecular clouds \cite{Balashev2017,Ranjan2018}; (iv) low impact parameter ESDLA with large H$_2$ column \cite{Ranjan2018}. 
In short, the study of H$_2$ absorbers at high-$z$ towards quasars and GRB afterglows opened a unique window to the understanding of the ISM properties in the overall galaxy population, similar to what has been done in the local ISM towards bright nearby stars.

\section{CUBES science case}
\label{sec:CUBES}
In spite of significant progress in recent years, the observations are now reaching a bottleneck, mostly due to the limiting instrumental capabilities to follow-up faint targets (typically $m_{\rm G}\gtrsim19$) that make the bulk of the new and most promising candidates. In fact, while observations were first obtained with UVES (since high resolution has a high priority for H$_2$ studies), the preference has then shifted to X-shooter to optimize the observing time for $m_{\rm G} \gtrsim 19$ targets. The drawback for choosing X-shooter is the loss in spectral resolution $R\sim6000$\footnote{H$_2$ absorption lines mostly falls in UVB arm of X-shooter}, which quite complicates the analysis and limits the derivation of physical quantities from the data (see e.g. \cite{Noterdaeme2017,Balashev2019} and comparison of the spectra in Fig.~\ref{fig:J1237}). Currently, most of the bright targets accessible from VLT have 
already been observed, and available candidates in SDSS are already out of reach for UVES and at the limit for X-shooter within reasonable observing time, see Fig.~\ref{fig:H2_cand}. 

CUBES \cite{CUBES} will exactly fill this gap, with a high efficiency in the blue, while keeping a high-resolution to maximise the scientific outcome. This is 
essential to continue investigating the cold neutral phase in and around galaxies at high redshifts. Indeed the efficiency of CUBES is about an order of magnitude higher than UVES, and the declared resolution ($R\sim20000$) is enough to resolve the H$_2$ lines much better than X-shooter currently can do. Using CUBES exposure time calculator \cite{CUBES_ETC}, we estimate that for a QSO with G-band magnitude around 20, $\rm S/N \approx 10$ can be reached in only one hour of observing time. This will allow one to increase the available samples by several fold, 
as needed for the relevant science goals. Additionally, the wavelength coverage of CUBES is already optimized for H$_2$ identified in SDSS (see Fig.~\ref{fig:H2_cand}).  In the following three subsections, we will briefly describe the most important CUBES science goals to our eyes, dealing with intervening and proximate H$_2$-bearing DLAs, i.e. cold gas in, respectively, the overall population (hence mostly faint) galaxies and the environment of massive active galaxies.

\subsection{Intervening DLAs: probing galaxies at high-z}
\label{sec:intervening}

\begin{figure*}
\centering
\includegraphics[trim={0.0cm 0.0cm 0.0cm 0.0cm},clip,width=1.0\textwidth]{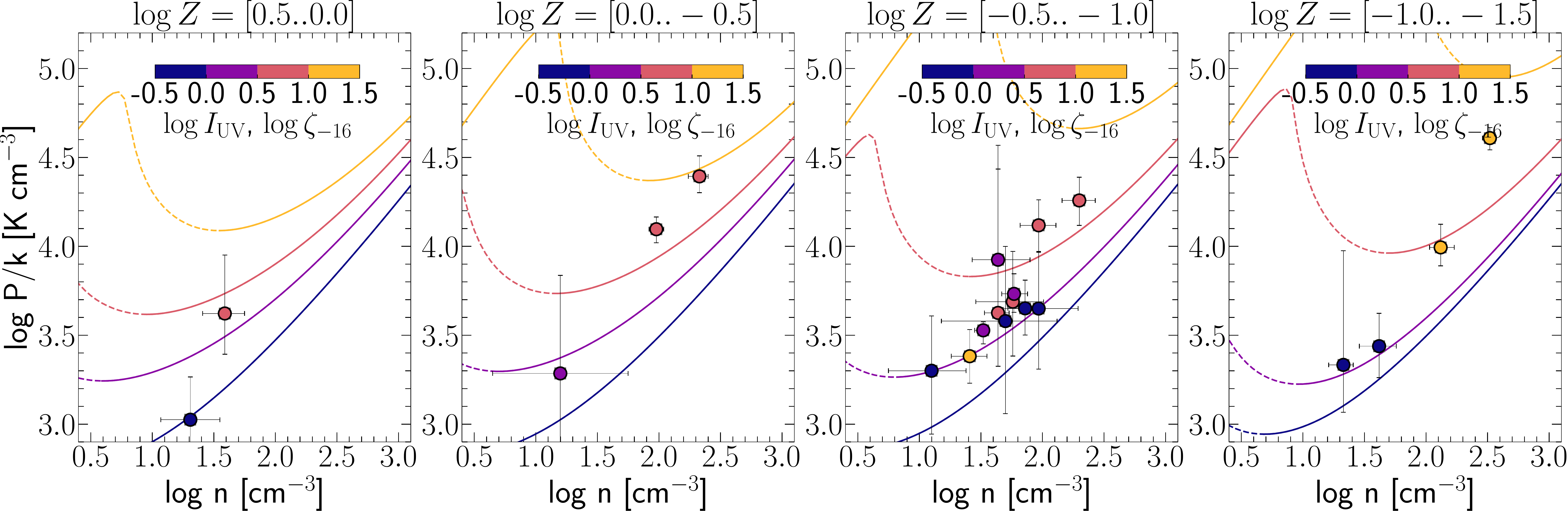}
\caption{The measured thermal pressure versus number density in intervening H$_2$ absorption systems at high redshifts. The  panels from left to right correspond to three samples of selected metallicities in $\log Z = 0.5, 0.0, -0.5, -1.0, -1.5$ bins, shown on the top of each panel. The curves on each plot show the calculation of the phase diagram at median metallicity of the bin. The color of the lines and symbols indicate a measured (and assumed) UV field (in Draine field units). For calculations we also co-scaled cosmic ray ionization rate with UV field (assuming that they are both associated with star formation), which is indicated by the color bar at the top of each panel. 
\label{fig:phase}}
\end{figure*}

While the first detected H$_2$ absorption system at high-$z$ was found at the quasar redshift 
(see recent re-analysis by \cite{Balashev2020a}), studies of H$_2$ initially focused on intervening systems -- absorption systems at cosmological remote distances from the quasar, i.e. that are not physically associated with the later. Typically, these are securely chosen by a velocity difference $\Delta v \approx (z_{\rm abs} - z_{\rm qso}) / (1 + z_{\rm qso}) c > \rm few\times 1000$\,km\,s$^{-1}$. This permits a large redshift path to be probed and therefore, such systems 
were preferentially selected in the first blind surveys.\footnote{However, since the cross section of associated systems are found to be much larger, as was found later, the incidence rate of intervening and proximate systems are not much different, see Section~\ref{sec:proximate}} 

The determination of the physical conditions was one of the primary goal of studies of intervening H$_2$ absorption systems
\cite{Srianand2005,Noterdaeme2007,Balashev2010}, since excitation of H$_2$ rotational levels (e.g. \cite{Noterdaeme2007b,Balashev2019}), as well as associated \CI\ fine-structure (e.g. \cite{Jorgenson2010,Balashev2017}), and rotational levels of HD (e.g. \cite{Balashev2010}) and CO molecules \cite{Srianand2008,Noterdaeme2011} provide a straight way to do it. The later two molecules were found in a handful of H$_2$-bearing DLAs systems, but provide valuable independent constraints on the physical conditions for molecular gas with high column, thanks to different sensitivities of the molecules on the various excitation processes. Indeed, the rotational levels of 
molecules as well as fine-structure levels of atomic species are populated by a competition between collisional excitation, radiative pumping (following the electronic UV transitions) and direct radiative excitation. Therefore the measured populations on these levels are sensitive to the rates of these processes and provide the way to constrain the kinetic temperature, number density, UV flux and Cosmic Microwave Background (CMB) temperature (mostly using CO, \cite{Noterdaeme2011}).

Originally, H$_2$-bearing DLAs were analysed mostly on a case-by-case basis. The main outcome of these studies were: (i) The measurement of the gas kinetic temperature, estimated using the ortho-to-para ratio\footnote{In the case of H$_2$, this corresponds to the ratio of J=1 and J=0 rotational levels.} of H$_2$ molecules, 
and giving $T_k \sim$100\,K, i.e. H$_2$ probe the cold phase of ISM; (ii) The measurement of the number densities, found in the range $n\approx10-\rm 300$\,cm$^{-3}$, i.e. also consistent with what is expected for the cold neutral ISM; (iii) The derivation of the strength of the UV field, found to span a large range from slightly lower than the Draine field up to two orders of magnitude higher. Recently performed systematic analyses of the emergent samples of H$_2$-bearing DLAs \cite{Klimenko2020,Kosenko2021} start to show interesting dependencies between estimated physical conditions. For example: (i) The thermal pressure measured in the cold phase of the ISM increases with increasing neutral gas column density \cite{Balashev2017,Balashev2019}. This is quite expected from analogy with local measurements as well as simulations, which shows that the high column density systems probe galaxies at the smallest impact parameters, and hence expected to have enhanced pressures; (ii) The ratio of UV field to number density correlates with the gas temperature \cite{Klimenko2020}. This correlation is also metallicity-dependent but it seems not to contradict with heating-cooling balance of the neutral ISM (see Fig.~\ref{fig:phase}). 
(iii) The cosmic ray ionization rate estimated using HD/H$_2$ ratio is found to depend quadratically on the strength of the UV field \cite{Kosenko2021}. Since cosmic ray ionization ray also determine the thermal balance especially at low metallicities \cite{Bialy2019}, this should also be take into account for future studies. To summarize, we anticipate many progresses in our understanding of the physics of the cold diffuse gas at high-$z$, and hence in our understanding of how star-formation occurs, when larger samples (probing wider ranges of the physical conditions and chemical enrichment) will get observed with CUBES.
 

\subsection{Proximate DLAs: probing AGNs outflowing gas}
\label{sec:proximate}

\begin{figure*}
\centering
\includegraphics[trim={0.0cm 0.0cm 0.0cm 0.0cm},clip,width=1.0\textwidth]{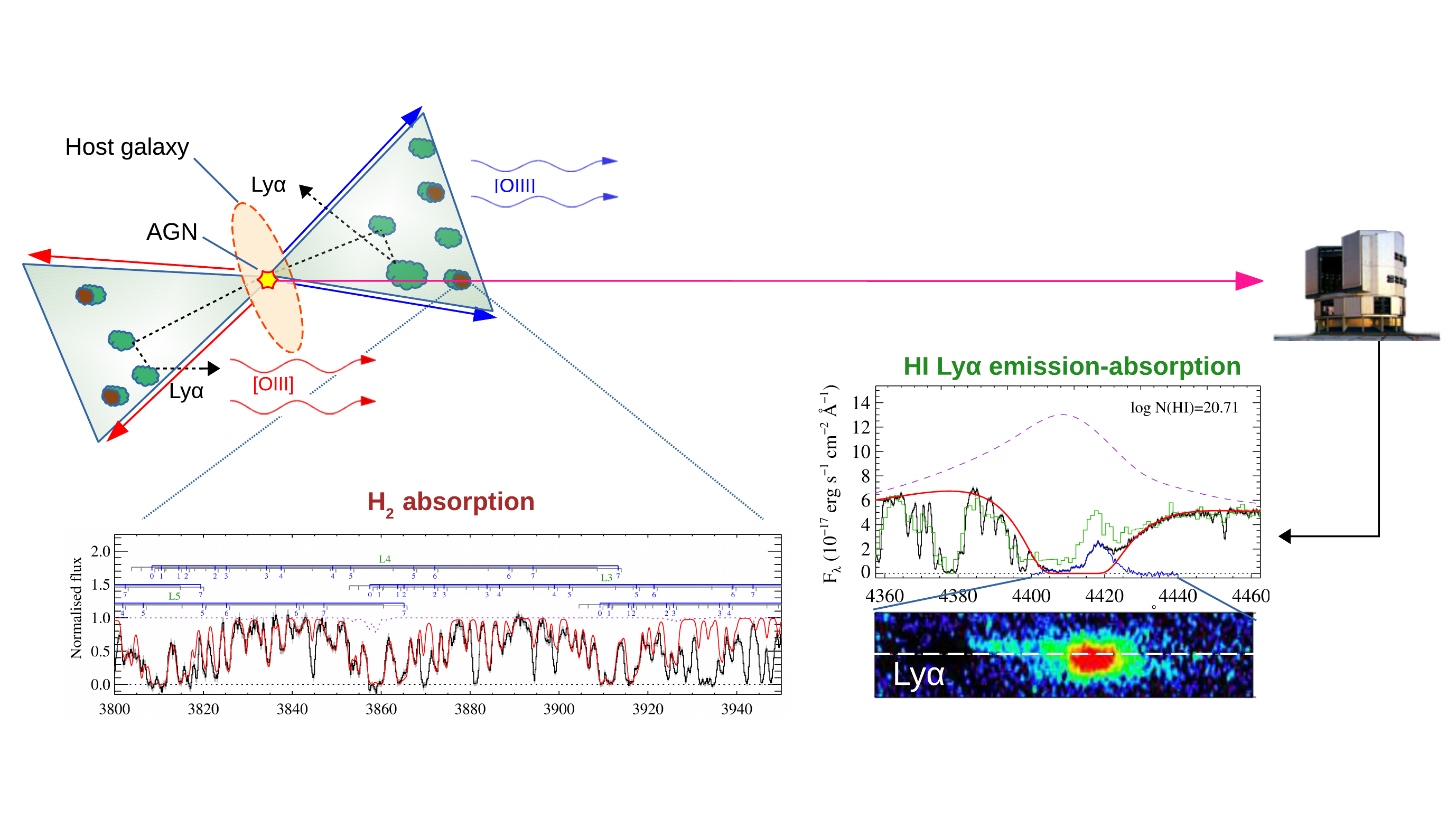}
\caption{Sketch representing studies of AGN outflowing gas using VLT/X-shooter observations (reproduced from \cite{Noterdaeme2021a}). The left and right bottom panels show portions of J\,$0015+1842$ spectrum covering H$_2$ absorption lines and Ly-$\alpha$ emission/absorption line, respectively. The detection of extended [O\,{\sc III}] and Ly-$\alpha$ emission (shown in 2d spectrum in the bottom panel) and surprisingly high excitation of H$_2$ rotational levels, indicate that this absorption system is associated with outflowing gas.
\label{fig:outflow}}
\end{figure*}

A recent study of the SDSS quasar database by \cite{Noterdaeme2019} uncovered a population of proximate (i.e. at sensibly the quasar redshift) H$_2$ absorption systems. It was found that the incidence rate of proximate H$_2$ is actually up to a factor of 10 higher than what is expected from the statistics of intervening ones.  
This is opposite to naive expectations that the strong radiation field of quasar should suppress the presence of H$_2$ in its vicinity, due to efficient photo-dissociation of the molecules. However, the observed strong excess hypothetically can be explained by either clustering of the matter and satellite galaxies around quasars or the presence of galactic-scale outflows, which, as current local observations indicate, frequently contains a large amount of molecular gas \cite{Veilleux2013,Cicone2014}. The detection of leaking Ly-$\alpha$ emission (seen as Ly-$\alpha$ emission inside the damped Ly-$\alpha$ absorption) in a roughly half of the candidates from SDSS \cite{Noterdaeme2019}, as well as follow-up observations \cite{Noterdaeme2021a} indicate that substantial fraction of proximate H$_2$ absorption systems may indeed arise from outflowing gas. In that sense, molecular hydrogen served as the species that allow the pre-selection of these systems in low-resolution spectra, but also provides a key diagnostic of the physical conditions in the diffuse molecular gas associated with outflow. For example, in the first and only one system of such class studied in details so far, J\,$0015+1842$ at $z=2.631$, it was found, using the excitation of H$_2$ rotational levels, that the number density is $n\sim10^4$\,cm$^{-3}$ and that the UV field is $\sim10^3$ times higher than the Draine field. This is significantly higher than what is typically measured in intervening DLAs. Assuming that the quasar produces most of the UV radiation, this permitted to constrain the distance between the AGN central engine and H$_2$-bearing medium to be $\sim 10-15$\,kpc, see Fig.~\ref{fig:outflow}. The analysis of J\,$0015+1842$ urges for 
observing a sample to enable 
statistical studies and 
address open questions about the nature of these systems, constraining key parameters of AGN outflows, such as the mass rate, fraction of active AGNs, opening angles etc. Additionally, the detection of such kind of systems opens wide opportunities for the multi-wavelength observations of AGN outflows at high-$z$, including near IR (to study H$_2$ in emission), sub-mm (to study cold molecular phase by CO emission \cite{Noterdaeme2021b} as well as other molecules) and makes valuable quantitative input for the modelling of outflowing gas.


\subsection{Probing dense gas}

As we already mentioned above, SDSS has, by construction, significant selection bias against dusty and red quasars \cite{Krogager2019}.  
Indeed, if dust is present anywhere along the line of sight, then the background quasar will appear fainter and redder than it intrinsically is (e.g. \cite{Pei1991,Pontzen2009}). Quasars with such dusty systems are therefore more likely to fall outside the color-selection and/or below the flux-limit of spectroscopic surveys.
Since, as local measurements indicate, the amount of dust obscuration scales with the H$_2$ column density (e.g. \cite{Savage1977,Roman-Duval2010,Shull2021}), this implies a bias against absorption systems with high H$_2$ column densities. These elusive systems correspond to translucent or dense molecular gas and therefore their detailed study is very important for 
a complete theory of the baryon cycle in the ISM over a wide range of metallicities. The next generation spectroscopic survey telescopes (such as 4-m Multi-Object Spectroscopic Telescope (4MOST), Dark Energy Spectroscopic Instrument, Mauna Kea Spectroscopic Explorer, William Hershel Telescope Enhanced Area Velocity Explorer, etc) will likely push further the quasar selection parameter space and obtain samples with significantly less selection bias against dusty absorbers. Follow-up observations of reddened quasars in the blue will naturally require a very good efficiency as can be brought by CUBES.  

\subsection{Astrophysical probes of fundamental physics}
Molecular hydrogen absorption systems also provide unique probes of fundamental physics and cosmology. By comparing the observed relative wavelengths of the numerous absorption lines from H$_2$ with those measured in the laboratory, it is possible to 
constrain the variation of the electron-to-proton mass ratio, $\mu$, over cosmological time scales \cite{Thompson1975,Varshalovich1993,Bagdonaite2012}, which gives typical limits of $|\Delta\mu / \mu|< 5 \times 10^{-6}$ \cite{Ubachs2016}. This value is limited by systematics (including instrumental ones, \cite{Rahmani2013}), the sensitivity of H$_2$ transitions to variation of $\mu$, as well as a lack of systems with suitable characteristics. 
While it was found that some molecules (e.g. CH$_3$OH, NH$_3$ and HC$_3$N) are much more sensitive to variation of $\mu$ \cite{Henkel2009,Kanekar2015}, these are very rarely detected in the distant Universe\footnote{This is due to combination of two factors: (i) the cross-section of dense molecular gas (that is needed to detect aforementioned complex molecules) is much smaller than that of  H$_2$-bearing gas (associated mostly with more diffuse cold medium) (ii) the small fraction of radio bright QSOs, which are needed for the analysis sensitive transition}. Therefore, H$_2$ will remain a main prove for variations of $\mu$. By surveying H$_2$, CUBES will bring new targets at different redshifts, which are essential to decrease statistical errors as well as alleviating a number of possible systematics related to the systems characteristics. 
Additionally, accompanying detections of HD and CO molecules in H$_2$ absorption systems can also provide independent constrain on $\mu$ \cite{Ivanov2008,Darpa2016,Dapra2017,Ubachs2019}. Moreover, as we mentioned above, the detection of CO associated with H$_2$ \cite{Srianand2008} provides a direct measurement of the CMB temperature at high redshift \cite{Noterdaeme2011}, a fundamental test for the adiabatic cooling of the Universe predicted by standard Big-Bang theory.

In short, while CUBES's design will not permit direct use of these astrophysical probes (because of a relatively short wavelength range and because it is not intended to have exquisite wavelength calibration), CUBES will be extremely effective in characterizing a large number of appropriate H$_2$-bearing DLAs, very much easing the selection for follow-up studies with higher resolution and stable spectrographs.

\subsection{Candidates}
\label{sec:cand}

\begin{figure*}
\includegraphics[width=1.0\textwidth]{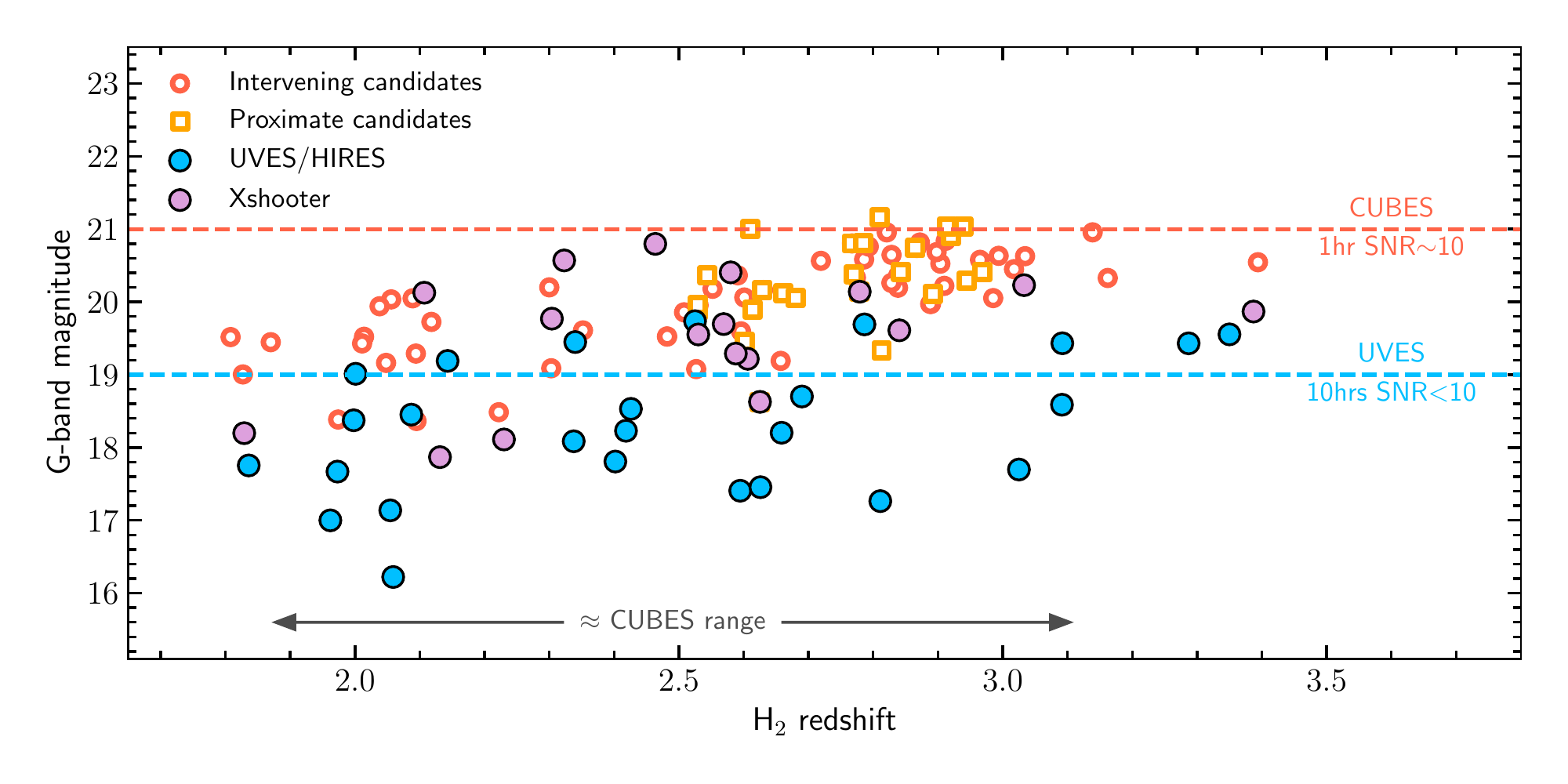}
\caption{G-band magnitude and  redshifts of H$_2$ absorption systems at high redshift. The blue and violet circles represent systems already confirmed and analysed using UVES/HIRES and X-shooter, respectively. The red and orange squares represent repectively the intervening and proximate H$_2$ candidates, pre-selected in the SDSS. The blue and red dashed lines indicate the typical magnitudes of the quasars that can be reached with UVES 10 hrs and CUBES 1hr exposure time, respectively, to achieve $\rm SNR\sim10$ in the spectrum. The redshift range accessible with CUBES is marked by the black horizontal solid lines with arrows.} 
\label{fig:H2_cand}
\end{figure*}

Fig.~\ref{fig:H2_cand} shows the G-band magnitude - redshift diagram for the confirmed (already followed-up) and candidates H$_2$ absorption systems (preselected in SDSS and that can be observed from Paranal). The intervening candidates were compiled based on DR14 using the direct-H$_2$ and \CI\ pre-selection techniques (for descriptions of techniques, see \cite{Ledoux2015,Balashev2014}) and proximate candidates were taken from \cite{Noterdaeme2019}.  The improved efficiency of the CUBES in comparison with UVES will allow to collect S/N$\sim 10$ spectra for all available candidates in less than one hour per object, when UVES requires already 10h for 2 magnitude fainter objects 
In addition, the wavelength range of CUBES corresponds to redshifts for H$_2$ lines that almost perfectly match the redshift distribution of the candidates. This shows that we already have a substantial number of the targets that can be straightforwardly follow-up with CUBES with high efficiency to multiply the available samples of intervening and proximate H$_2$-bearing DLAs.

\section{Conclusions}
\label{sec:conclusion}

Investigating molecular hydrogen absorption systems at high-z provides an extremely powerful probe of the physical conditions of the cold phase of the interstellar medium in the overall population of remote galaxies. The stepped-like progress in these studies has always been determined by observational capabilities, where two evident boosts were associated with ($i$) the advent of 8-m class telescopes (KECK and VLT) and $(ii)$ the development of efficient pre-selection techniques in massive spectroscopic surveys (SDSS). We are now facing a bottleneck since follow-up studies of the many targets identified in the SDSS are faint and already reach the capabilities of current instruments. This issue can be overcome with CUBES - the next generation efficient blue spectrograph that is planned to be mounted on the VLT. We briefly discussed three H$_2$-related top-priority science cases for CUBES in our opinion: 
($i$) Studies of the thermal state of the cold ISM in the wide range of the environments sampled by intervening DLAs, (ii) Detailed investigations of the molecular gas in AGN outflows at high redshifts (iii) Probes of translucent and dark molecular gas in absorption via hunting for the dusty and high-column H$_2$-bearing DLAs that can be exclusively be reached with CUBES. This, together with the forthcoming 
generation of spectroscopic surveys, such as 4MOST (which will provide better exploration of the sky accessible from Paranal) will definitely provide the next boost in understanding the small-scale physics in the gas over a wide range of redshifts and environments. 

\section*{Conflict of interests}

The authors declare that they have no conflict of interest.

\section*{Data availability}

The data underlying this article will be shared on reasonable request
to the corresponding author.

\begin{acknowledgements}
SB is supported by Russian Science Foundation grant 18-12-00301.
PN is supported by the French {\sl Agence Nationale de la Recherche under grant ANR-17-CE31-0011 (``HIH2'')}. 
\end{acknowledgements}

%
%

\bibliographystyle{spmpsci}      
\bibliography{references.bib}   


\end{document}